\documentclass[journal]{IEEEtran}
\IEEEoverridecommandlockouts
\usepackage{lmodern,textcomp}
\usepackage{graphics}
\usepackage{multicol}
\usepackage{lipsum}
\usepackage{color}
\usepackage{amsthm}
\usepackage[demo]{graphicx}
\usepackage{amssymb}
\usepackage{subfig}
\usepackage{mathtools}
\usepackage{amsbsy}
\usepackage{setspace}
\usepackage{float}
\usepackage{lettrine}
\usepackage{newclude}
\usepackage[normalem]{ulem}
\usepackage{latexsym}
\usepackage{multirow}
\usepackage{rotating}
\usepackage{booktabs}
\usepackage{wasysym}
\usepackage{mathrsfs}
\usepackage{bbm} 
\usepackage{epsfig,cite,amsfonts,psfrag}
\usepackage{apptools}
\usepackage{graphicx}
\usepackage[table]{xcolor}
\usepackage{chngcntr}
\usepackage{blindtext}
\usepackage{hyperref}
\usepackage{optidef}
\usepackage{accents}
\usepackage{dsfont}
\usepackage{siunitx}
\usepackage{overpic}
\usepackage{amsmath,amssymb,amsfonts}
\usepackage{graphicx}
\usepackage{textcomp}
\usepackage{xcolor}
\usepackage{optidef}
\usepackage{lipsum}
\usepackage{amsthm}
\usepackage{array}
\usepackage{color}
\usepackage{url}
 
\usepackage{optidef}
\usepackage{times}
\usepackage{fancyhdr,graphicx,amsmath,amssymb}
\usepackage[ruled,linesnumbered,noend]{algorithm2e}
\usepackage{algorithmicx}
\usepackage{algcompatible}
\IEEEoverridecommandlockouts
\usepackage{cite}
\usepackage{graphicx}
\usepackage{url}
\usepackage{relsize}
\usepackage{hhline}
\usepackage{nicefrac}
\usepackage{amssymb,bm,upgreek}
\usepackage[table]{xcolor}
\usepackage{multirow}

\theoremstyle{plain}

\newtheorem{remark}{Remark}

\newcommand{\vect}[1]{\mathbf{#1}}
\newcommand{\maximize}[1]{{\underset{{#1}}{\mathrm{maximize}}}}
\newcommand{\minimize}[1]{{\underset{{#1}}{\mathrm{minimize}}}}

\newcommand{\bl}[1]{\boldsymbol{#1}}

\def\BibTeX{{\rm B\kern-.05em{\sc i\kern-.025em b}\kern-.08em
    T\kern-.1667em\lower.7ex\hbox{E}\kern-.125emX}}

\def\diag{\mathrm{diag}}

\begin{document}

\title{Splitting Precoding with Subspace Selection and Quantized Refinement for Massive MIMO}

\author{\IEEEauthorblockN{\normalsize
    Yasaman Khorsandmanesh, \textit{Student Member, IEEE},  Emil Björnson, \textit{Fellow, IEEE},  \\ and Joakim Jaldén, \textit{Senior Member, IEEE}  }
    \thanks{
Y.\ Khorsandmanesh, E.\ Björnson, and J.\ Jaldén are with the School of Electrical Engineering and Computer Science, KTH Royal Institute of Technology, Stockholm, Sweden (E-mails: \{yasamank, emilbjo, jalden\}@kth.se).}
\thanks{This work was supported by the Knut and Alice Wallenberg Foundation and Vinnova through the SweWIN center (2023-00572).}
}

\maketitle

\begin{abstract}
Limited fronthaul capacity is a practical bottleneck in massive multiple-input multiple-output (MIMO) 5G architectures, where a base station (BS) consists of an advanced antenna system (AAS) connected to a baseband unit (BBU). Conventional downlink designs place the entire precoding computation at the BBU and transmit a high-dimensional precoding matrix over the fronthaul, resulting in substantial quantization losses and signaling overhead. This letter proposes a splitting precoding architecture that separates the design between the AAS and BBU. The AAS performs a local subspace selection to reduce the channel dimensionality, while the BBU computes an optimized quantized refinement precoding based on the resulting effective channel. 
The numerical results show that the proposed splitting precoding strategy achieves higher sum spectral efficiency than conventional one-stage precoding.
\end{abstract}

\vspace{-3mm}

\section{Introduction}
{
Massive multiple-input multiple-output (MIMO) significantly improves spectral efficiency (SE) compared to traditional single-antenna systems by serving many user equipments (UEs) simultaneously through spatial multiplexing~\cite{bjornson2017massive}.} In practical 5G deployments, the base station (BS) typically consists of a centralized baseband unit (BBU) and a distributed advanced antenna system (AAS) connected via a fronthaul link with limited capacity. The BBU  performs digital baseband processing related to the received uplink data and transmitted downlink data. The AAS is a box that integrates the antenna elements and radio units. This splitting architecture is aligned with the open radio access network (O-RAN) principles \cite{peng2015fronthaul}.

As the number of antennas at the AAS increases, conventional fronthaul transmission becomes a major bottleneck. Both the UE's data symbols and the precoding matrix must be transferred over this fronthaul, and the limited link capacity requires quantization. Transmitting precoded signals at high resolution is therefore infeasible in massive MIMO systems. This motivates the development of alternative architectures that reduce fronthaul load while maintaining system performance.

{Previous research has investigated finite-resolution precoding, where the precoded signals are quantized at the BBU prior to fronthaul transmission~\cite{castaneda2019finite}. However, in massive MIMO systems, this approach imposes substantial fronthaul overhead. A more efficient approach is proposed in~\cite{khorsandmanesh2023optimized}, where a quantization-aware precoding method tailored to fronthaul-constrained systems is developed. Instead of quantizing the precoded signal, this transfers the data without quantization and focuses on optimizing a quantized precoding matrix through a mixed-integer formulation that minimizes the sum mean-squared error (MSE). While suitable for small-scale systems, such single-stage precoding designs become impractical in large-scale settings due to the high computational overhead and the cost of transmitting full-dimensional precoding matrices over the fronthaul, which also limits architectural flexibility. With recent advances in distributed processing, the AAS can now perform part of the physical-layer computation locally~\cite{karthikeyan2025advanced}. For example, \cite{chang2016prefiltering} proposes a pre-filtering architecture where the AAS that compresses the fronthaul data dimensionality, achieving near-lossless performance under perfect channel state information (CSI); however, this work focuses on channel estimation rather than precoding. Two-stage precoding structures were explored in hybrid beamforming~\cite{sohrabi2016hybrid}, but remain largely unexplored in fully-digital downlink systems with fronthaul constraints. A related idea is the two-stage reduced-dimensionality codebook in~\cite{almradi2020two}, which uses statistical CSI for analog beamforming and quantized effective channel for digital refinement. Moreover, \cite{kang2016layered} proposes a layered precoding strategy based on long-term channel statistics for the central-RAN (C-RAN) architecture.

In this letter, we propose distributing computation across the AAS and BBU. The AAS sets a precoding that captures the \emph{dominant signal directions} and provides a high-gain, low-dimensional effective channel, whereas the BBU designs a \emph{quantized} precoding for suppressing multi-user interference. This reduces fronthaul load while keeping precoding flexibility at the AAS side. The main contributions are:
\vspace{-1mm}
\begin{itemize}
\item We propose a \emph{splitting} precoding architecture for fronthaul-limited massive MIMO, in which the AAS performs subspace selection through a semi-unitary precoding, while the BBU computes an optimal quantized refinement precoding that minimizes the MSE.
\item To maximize the signal gain, we consider maximum-ratio transmission (MRT) based subspace selection at the AAS, including its Gram–Schmidt orthogonalized variant, and also explore a discrete Fourier transform (DFT) precoding as a practical low-complexity alternative.
\item The BBU quantization-aware precoding is obtained by solving an MSE minimization problem under quantization constraints and
using a Schnorr–Euchner sphere decoding (SESD)–based algorithm.
\item We provide numerical results that demonstrate the benefits of the proposed splitting design over the single-stage quantization-aware baseline with the same fronthaul resolution, particularly at high signal-to-noise ratio (SNR).
\end{itemize}}
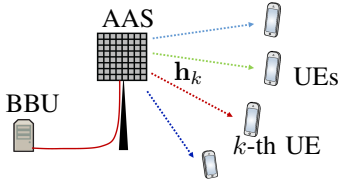
\begin{figure}[t!]
  \centering
   \begin{overpic}[width=0.4\linewidth]{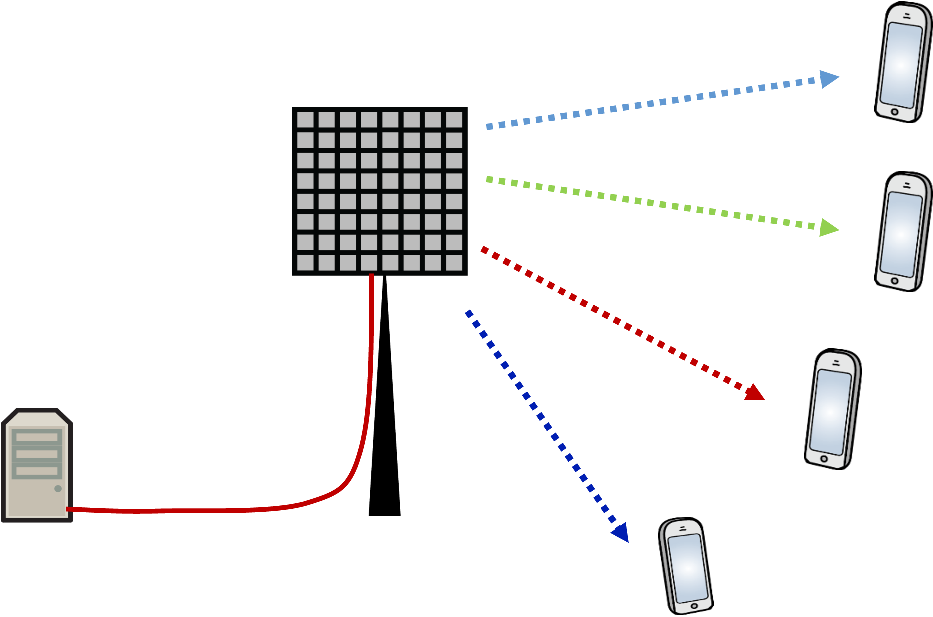}
  \put(-3,25){BBU}%
  \put(33,57){AAS}%
  \put(105,35){UEs}%
  \put(82,9){$k$-th UE}%
  \put(60,37){$\mathbf{h}_k$}%
   \end{overpic}
\caption{The considered downlink massive MIMO system with splitting precoding and fronthaul constraint.
}
\label{fig:systemmodel}
\vspace{-5mm}
\end{figure}
\section{System Model}
We consider the downlink of a single-cell massive MIMO system in which the BS transmits data to $K$ single-antenna UEs over a common time-frequency resource block. The BS comprises two key components: AAS, which consists of large $M \ge K$ antenna-integrated radios, and the BBU, connected via a fronthaul link with limited capacity, as depicted in Fig.~\ref{fig:systemmodel}. Any signals exchanged between the BBU and the AAS must be quantized to finite resolution.
In the following, the terms \emph{precoding} and \emph{precoding matrix} are used interchangeably.

Since the data symbols are drawn from a channel coding codebook, they can be transmitted over the fronthaul link without quantization errors. The modulation mapping is performed at the AAS. In contrast, the precoding matrix consists of complex-valued entries and must be quantized prior to fronthaul transmission. This matrix remains constant over a channel coherence interval. To focus on the algorithmic development of splitting precoding, we assume that the AAS and the BBU have perfect CSI. 
\vspace{-3mm}
\subsection{Downlink Transmission with Splitting Precoding} \label{sec2chamodel}
The downlink system model is given by
\begin{equation}
\mathbf{y} = \mathbf{H}  \mathbf{P}_{\rm A} \mathbf{P}_{\rm B} \mathbf{s} + \mathbf{n},
\end{equation}
where $\mathbf{y} = [y_1, \ldots, y_K]^\mathrm{T} \in \mathbb{C}^{K}$ contains the received signals at all the UEs, and $y_k $ denotes the signal received at the $k$-th UE. The downlink channel matrix $\mathbf{H}= [\mathbf{h}_1, \ldots, \mathbf{h}_K]^\mathrm{T} \in \mathbb{C}^{K \times M}$ is a flat-fading narrowband channel. The vector $\mathbf{n} = [n_1, \ldots, n_K]^\mathrm{T} \in \mathbb{C}^{K}$ is the additive white Gaussian noise, with i.i.d. entries $n_k \sim \mathcal{CN}(0, \sigma_0^2)$. Here, $\mathbf{s} = [s_1, \ldots, s_K]^\mathrm{T} \in \mathcal{O}^{K}$ contains the data symbols with $s_k$ denoting the unit-power symbol intended for UE $k$, and $\mathcal{O}$ denotes a finite constellation set such as QAM. The AAS precoding $\mathbf{P}_{\rm A} \in \mathbb{C}^{M \times N}$ is computed locally at the AAS and performs subspace selection. Based on the available effective channel (channel times AAS precoding), the BBU computes the BBU precoding $\mathbf{P}_{\rm B} \in \mathcal{P}^{N \times K}$ in a lower-dimensional subspace and quantizes it before transmission over the fronthaul. This split structure reduces the fronthaul load, since $\mathbf{P}_{\rm B}$ has lower dimensionality than a full $M \times K$ precoding matrix, allowing higher quantization resolution to be used for a given fronthaul capacity.

The quantization alphabet $\mathcal{P}$ is a discrete subset of $\mathbb{C}$:
\begin{equation}
   \mathcal{P} = \{ l_{R} + j l_{I} : l_{R}, l_{I} \in \mathcal{L} \},
   \label{eq:quantizationset2stage}
\end{equation}
where $\mathcal{L} = \{ l_0, \ldots, l_{L-1} \}$ contains $L$ real-valued quantization levels.  The number of quantization bits per real dimension is $B = \log_2 (L)$. Uniform quantization is assumed for both real and imaginary parts, and details are provided in Sec.~\ref{sec2quantize}. For brevity, we only consider a rectangular quantization grid, but the proposed method is easily extendable to a polar grid.

\begin{remark}
Assume the total fronthaul link budget is $C_{\rm Total}$. With the proposed bit-splitting architecture, the fronthaul load scales as $2 B_{\rm Split} N K$, whereas a conventional one-stage design requires $2 B_{\rm One\text{-}stage} M K$. Since $K \le N < M$, the proposed architecture incurs a lower fronthaul load and can therefore support a higher quantization resolution. For a fixed total fronthaul budget $C_{\rm Total}$, the corresponding bit allocations satisfy $\frac{M}{N} = \frac{B_{\rm Split}}{B_{\rm One\text{-}stage}}$.
\end{remark}

The AAS computes the transmit vector $\mathbf{x} = \mathbf{P}_{\rm A} \mathbf{P}_{\rm B} \mathbf{s}$, and the average transmit power must satisfy
\begin{equation}
    \mathbb{E}\left[\| \mathbf{x} \|_2^2\right] = \left\| \mathbf{P}_{\rm A} \mathbf{P}_{\rm B} \right\|_{\mathrm{F}}^2 \le q,
    \label{eq:powercon2stage}
\end{equation}
where $\| \cdot \|_{\mathrm{F}}$ denotes the Frobenius norm, and $q$ is the maximum average transmit power. The equality holds due to the unit power assumption on the data symbols, i.e., $\mathbb{E}[\mathbf{s} \mathbf{s}^H] = \mathbf{I}_K$. {  We propose to select $\mathbf{P}_{\rm A}$ as a semi-unitary matrix, in which case the power constraint reduces to $\| \mathbf{P}_{\rm B}\|_{\mathrm{F}} ^ 2 \leq q$. }

The resulting precoded signal vector is transmitted wirelessly over the downlink channel towards the UEs. Each UE estimates its intended data symbol using linear equalization as $\hat{s}_k = \beta_k y_k$, where $\beta_k \in \mathbb{C}$ is a complex receiver gain selected to minimize the MSE $\mathbb{E}[|s_k - \hat{s}_k|^2]$. The system design aims to treat all UEs fairly by minimizing the total sum MSE:
\begin{equation}
\mathbb{E}\left[ \| \mathbf{s} - \hat{\mathbf{s}} \|_2^2 \right] = \mathbb{E}\left[ \| \mathbf{s} - \mathbf{B} \mathbf{y} \|_2^2 \right],
\label{eq:mmseform2stage}
\end{equation}
where $\hat{\mathbf{s}} = [\hat{s}_1, \ldots, \hat{s}_K]^\mathrm{T}$ contains the estimated data symbols and $\mathbf{B} = \mathrm{diag}(\boldsymbol{\beta})$ is a diagonal matrix collecting the receiver gains $\boldsymbol{\beta} = [\beta_1, \ldots, \beta_K]^\mathrm{T}$, which may be optimized jointly with the precoding matrices.

Classical methods for computing continuous one-stage precoding include zero-forcing (ZF), regularized ZF (RZF), and MRT \cite{bjornson2014beamforming}. Among these, RZF precoding is the most desirable, as it minimizes the MSE in \eqref{eq:mmseform2stage}, but this approach is not optimal in the presence of quantization. To account for fronthaul constraints, we adopt a quantized RZF (QRZF) design, where the corresponding unnormalized continuous precoding matrix is given by
\begin{equation}
\bar{\mathbf{P}}^{\mathrm{QRZF}} =     \mathbf{H}^\mathrm{H} \left(\mathbf{H}\mathbf{H}^\mathrm{H}+ \mu\mathbf{I}_K \right)^{-1},
    \label{eq:wf}
\end{equation}
where $\mu = \frac{K}{(1-\eta)^2} \Big( \frac{\sigma_0^2}{q} + \eta(1-\eta)M\Big)$ is the quantized regularization factor and $\eta$ is { the distortion factor that depends on the quantization resolution \cite{fan2015uplink}.}
\vspace{-3mm}
\subsection{Offline Uniform Quantizer Function} \label{sec2quantize}
We adopt a \emph{data-driven} approach in which the step size $\Delta$ of a symmetric mid-rise uniform quantizer is optimized directly from the realizations of the unquantized BBU precoding $\bar{\mathbf{P}}_{\rm B}$ by minimizing the expected MSE distortion of the real and imaginary entries of the unquantized BBU precoding. Let $r$ denote a generic precoding entry and $\mathbb{Q}(\cdot)$ a real-valued quantizer. Under the maximum-entropy assumption, the optimal step size solves
\begin{equation}
    \Delta^\star = \arg\min_{\Delta>0}
    \mathbb{E}\!\left[\, |r - \mathbb{Q}(r)|^2 \,\right].
    \label{eq:delta_opt_expectation}
\end{equation}
Since this expectation is not available in closed form, it is approximated in practice by a sample mean computed from a set of independently generated BBU precoding realizations.

For a real value $x$, the $L$-level mid-rise quantizer is defined as
\begin{align}
    &
    \mathbb{Q}(x)
    = \Delta \left( o(x) - \frac{L-1}{2} \right), \\
&\quad \text{for } \quad
    o(x)
    = \min\!\left\{
        \max\!\left\{
            \left\lfloor \frac{x}{\Delta} + \frac{L}{2} \right\rfloor,\, 0
        \right\},\,
        L - 1
    \right\},
\end{align}
and the minimizing $\Delta^\star$ is obtained via a simple one-dimensional search. After this offline optimization, the quantizer remains fixed for all subsequent precoding matrices.

\vspace{-2mm}
\subsection{Problem Formulation}
We propose to jointly optimize the AAS precoding $\mathbf{P}_{\rm A}$, the quantized BBU precoding $\mathbf{P}_{\rm B}$, and the receiver scaling vector $\beta$. Since interference suppression is handled by the BBU stage, the AAS aims to select an $N$-dimensional subspace that maximizes the desired signal gain. Moreover, as the quantized BBU precoding cannot fully suppress multi-user interference, the AAS can further improve performance by incorporating interference mitigation already at the subspace selection stage. This leads to the following problem:
\begin{equation} 
\begin{aligned} 
\label{eq:AAS-PRECODING} \maximize{\mathbf{P}_{\rm A} \in \mathbb{C}^{M \times N}}\,\, &\big\lVert \mathbf{H}\mathbf{P}_{\rm A} \big \rVert_{\rm F}^{2} \\  \mathrm{subject~ to}\quad & \mathbf{P}_{\rm A}^{\mathrm{H}}\mathbf{P}_{\rm A} = \mathbf{I}_{N}.
\end{aligned} 
\end{equation}
To suppress residual interference inside the AAS-selected subspace, the BBU designs a quantized precoding that minimizes the sum MSE under the total power constraint in~\eqref{eq:powercon2stage}, explicitly accounting for the fronthaul quantization-aware $\mathbf{P}_{\rm B}$.\footnote{For sum-rate maximization, one may use the standard equivalence to weighted sum-MSE minimization; hence, we adopt the sum-MSE criterion as the central design objective.}
The corresponding optimization problem is
\begin{equation} \begin{aligned} \label{eq:main_problem} \minimize{ \bl{\beta} \in \mathbb{C}^K, \mathbf{P}_{\rm B} \in \mathcal{P}^{N \times K}}\,\, &\mathbb{E} \left \{ \| \vect{s} - \mathbf{B} \vect{y}\|^2\right \} \\ \mathrm{subject~ to}\quad & \| \mathbf{P}_{\rm B}\|_{\mathrm{F}} ^ 2 \leq q,
\end{aligned} 
\end{equation}
where $\mathbf{B}=\diag(\bl{\beta})$ is the diagonal matrix with the $k$-th diagonal entry, denoted $\beta_k$.
\section{Proposed splitting precoding}
In the following, we propose different approaches for the selection of the AAS and BBU precoding.
\subsection{AAS Precoding}
The AAS selects the subspace that maximizes the received signal strength. 

\subsubsection{Method I-- Gram--Schmidt MRT}
When the AAS precoding dimension satisfies $N=K$,  we construct an orthonormal basis from the MRT directions as a solution to \eqref{eq:AAS-PRECODING}. Let $\mathbf{h}_i$ denote the $i$th column of the downlink channel matrix $\mathbf{H}$. The AAS precoding is obtained by applying Gram--Schmidt orthogonalization to the MRT vectors:
\begin{equation}
    \mathbf{P}_{\rm A}^{\mathrm{GS\text{-}MRT}}
    = \mathrm{GS}\big( \mathbf{h}_1^{*}, \ldots, \mathbf{h}_N^{*} \big),
\end{equation}
where $\mathrm{GS}(\cdot)$ denotes the classical Gram--Schmidt procedure \cite{matsumura2011orthogonal}. The resulting semi-unitary basis decorrelates the MRT vectors and improves the BBU refinement by mitigating interference. While one could obtain a similar orthogonal subspace by selecting the dominant right singular vectors of the channel matrix (via an instantaneous singular value decomposition (SVD)), GS-MRT provides identical performance at significantly lower computational complexity.

\subsubsection{Method II-- MRT}
A low-complexity solution to maximize signal gain is obtained by using normalized MRT. The resulting AAS precoding is
\begin{equation}
    \mathbf{P}_{\rm A}^{\mathrm{MRT}}
    = \bigg[
        \frac{\mathbf{h}_1^{*}}{\|\mathbf{h}_1\|},
        \, \ldots,\,
        \frac{\mathbf{h}_N^{*}}{\|\mathbf{h}_N\|}
      \bigg],
\end{equation}
which is a computationally efficient approximation to \eqref{eq:AAS-PRECODING} without satisfying constraint $\mathbf{P}_{\rm A}^{\mathrm{H}}\mathbf{P}_{\rm A} = \mathbf{I}_{N}$ \footnote{We note that this approach may potentially violate the power constraint in \eqref{eq:powercon2stage}; however, to ensure a fair comparison among the algorithms, the final precoding matrix  is always scaled to satisfy the power constraint with equality.}.

\subsubsection{Method III-- DFT-Based Selection}
The previous methods require instantaneous channel processing at the AAS. To further reduce complexity, the AAS could transform the channel to the angular beamspace using a DFT matrix $\mathbf{F} \in \mathbb{C}^{M \times M}$ 
\begin{equation}
    \widetilde{\mathbf{H}} = \mathbf{H}\mathbf{F}.
\end{equation}
We then select the $N$ DFT columns that result in the columns of $\widetilde{\mathbf{H}}$ with the largest $\ell_2$-norms, which capture the dominant angular directions of the UEs. 
This method provides a semi-unitary subspace with minimal computational burden \cite{el2014spatially}, but might require $N$ to be larger than $K$ to achieve good performance. 
\vspace{-4mm}
\begin{figure*}
\begin{mini}|l|
{\mathbf{a}_i\in {\mathcal{P}}^{N \times 1}, i=1,\ldots,K }{\sum_{i=1}^K \Big( \mathbf{a}_i^\mathrm{H} \left ( \mathbf{H}_{\rm eff}^\mathrm{H}\mathbf{B}^\mathrm{H} \mathbf{B}\mathbf{H}_{\rm eff}+\lambda \mathbf{I}_N \right )\mathbf{a}_i -\mathbf{h}_{{\rm eff}_i}^\mathrm{T}\mathbf{a}_i - \left ( \mathbf{h}_{{\rm eff}_i}^\mathrm{T}\mathbf{a}_i\right )^\mathrm{H}\Big)}{}{}
 \label{eq:modifysd}
\end{mini}
\hrule
\end{figure*}
\subsection{BBU Precoding}
Problem \eqref{eq:main_problem} is non-convex because the optimization variables are coupled in the objective function, and one of the variables is discrete. To solve this problem, we approach it using the coordinate descent algorithm.

We let the effective channel as $\mathbf{H}_{\rm eff}= \mathbf{H} \mathbf{P}_{\rm A}$. Taking a Lagrangian approach, the optimization problem in \eqref{eq:main_problem} is written as 
\begin{align}
  &\mathfrak{L}(\mathbf{P}, \boldsymbol \beta, \lambda) = \lambda \big( \mathrm{tr}(\mathbf{P}_{\rm B}\mathbf{P}_{\rm B}^\mathrm{H})-q \big) \nonumber \\
  &+\mathrm{tr} \Big( \mathbf{I}_K-\mathbf{B} \mathbf{H}_{\rm eff}\mathbf{P}_{\rm B}-  \mathbf{P}_{\rm B}^\mathrm{H}\mathbf{H}_{\rm eff}^\mathrm{H}\mathbf{B}^\mathrm{H} + \mathbf{B}\mathbf{H}_{\rm eff}\mathbf{P}_{\rm B}\mathbf{P}_{\rm B}^\mathrm{H}\mathbf{H}_{\rm eff}^\mathrm{H}\mathbf{B}^\mathrm{H} \Big),
\end{align}
where $\lambda$ is the Lagrange multiplier. Fixing $\mathbf{P}_{\rm B}$ and $\mathbf{P}_{\rm A}$, the optimal receiver gain $\beta_k$ is obtained by computing the Wirtinger derivative and is given by
\begin{equation}
    \beta_k^{\mathrm{opt}} = \frac{ [\mathbf{P}_{\rm B}^H \mathbf{H}_{\rm eff}^H]_{k,k} }{ [\mathbf{P}_{\rm B}^H \mathbf{H}_{\rm eff}^H \mathbf{H}_{\rm eff} \mathbf{P}_{\rm B}]_{k,k} + \sigma_0^2 }.
\end{equation}
Since iterative joint optimization is computationally heavy, $\beta$ is fixed with  $\mathbf{P}$ set to the QRZF solution in \eqref{eq:wf}. For a fixed value of $\lambda$, the remaining problem is an integer convex program over the quantized search space of $\mathcal{P}$. While exact solvers (e.g., CVX with Gurobi) are feasible for moderately sized systems, an efficient alternative is to reformulate the problem as an integer least-squares problem and solve it using the SESD algorithm, which performs a structured tree search over the quantization points and prunes suboptimal branches to achieve the optimal MSE at substantially lower complexity.

By following the same steps as in \cite{khorsandmanesh2023optimized}, we can obtain \eqref{eq:modifysd} on the top of this page, where  $\mathbf{a} = \mathrm{vec}(\mathbf{P}_{\rm B})$ and $\mathbf{h}_{\rm eff} = \mathrm{vec}\Big( (\mathbf{B}\mathbf{H}_{\rm eff})^\mathrm{T}\Big)$.  By defining $\mathbf{V} = \mathbf{H}_{\rm eff}^\mathrm{H}\mathbf{B}^\mathrm{H} \mathbf{B}\mathbf{H}_{\rm eff}+\lambda \mathbf{I}_N$, we obtain the equivalent formulation of each term of the objective function in \eqref{eq:modifysd} as 
\begin{align}
    \mathbf{a}_i^\mathrm{H} \mathbf{V} \mathbf{a}_i -\mathbf{h}_{{\rm eff}_i}^\mathrm{T}\mathbf{a}_i - \left ( \mathbf{h}_{{\rm eff}_i}^\mathrm{T}\mathbf{a}_i\right )^\mathrm{H} = \lVert \mathbf{e}_i - \mathbf{R}\mathbf{a}_i \rVert_2^2 - \mathbf{e}_i^\mathrm{H}\mathbf{e}_i,
\end{align}
where $\mathbf{R} \in \mathbb{C}^{N \times N}$ is obtained from the Cholesky decomposition $\mathbf{V} = \mathbf{R}^\mathrm{H} \mathbf{R}$ and $\mathbf{e}_i = (\mathbf{h}_{{\rm eff}_i}^\mathrm{T}\mathbf{R}^{-1})^\mathrm{H}$. As $\mathbf{e}_i \in \mathbb{C}^{N \times 1 }$ does not depend on the optimization variable, we can rewrite the subproblem with respect to $\mathbf{a}_i$ as 
\begin{mini}|l|
	  {\mathbf{a}_i\in {\mathcal{P}}^{N \times 1} }{\lVert \mathbf{e}_i - \mathbf{R}\mathbf{a}_i \rVert_2^2.}{}{}
 \label{eq:sdlasteq1}
\end{mini}
Now, for a fixed value of $\lambda$, this minimization problem can be solved using the SESD algorithm. We then make a bisection search \cite{burden19852} over $\lambda$ to find the value that gives a solution that satisfies the power constraint $\| \mathbf{P}_{\rm B}\|_{\mathrm{F}} ^ 2 \leq q$ near equality.
\begin{figure}[!t] 
    \centering
    \subfloat[i.i.d. Rayleigh fading channel with $M=32$ and $K=8$.]{\label{fig:compare}{\includegraphics[width=0.4\textwidth]{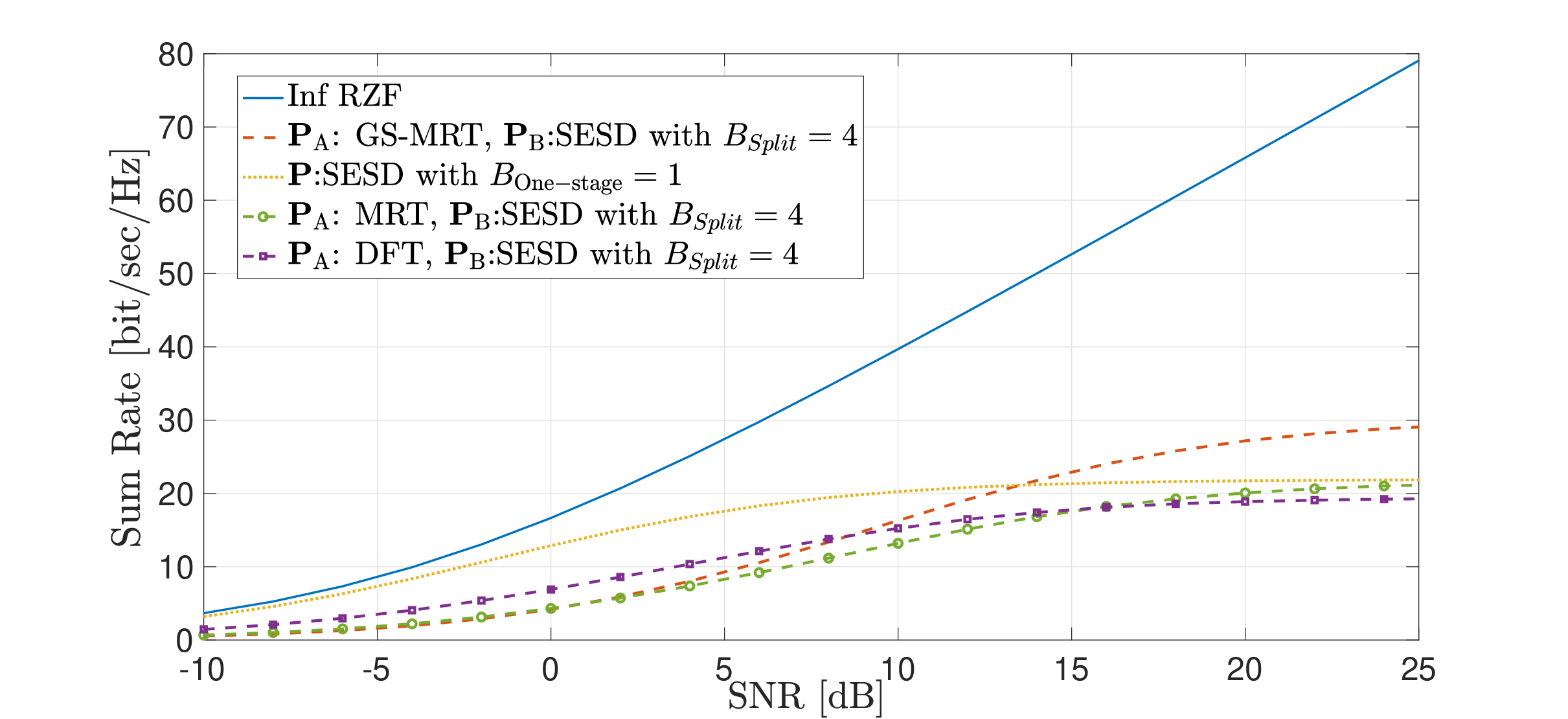} }}%
    \quad
 \subfloat[mmWave channel with $M=128$ and $K=8$.]{\label{fig:mmwave}{\includegraphics[width=0.4\textwidth]{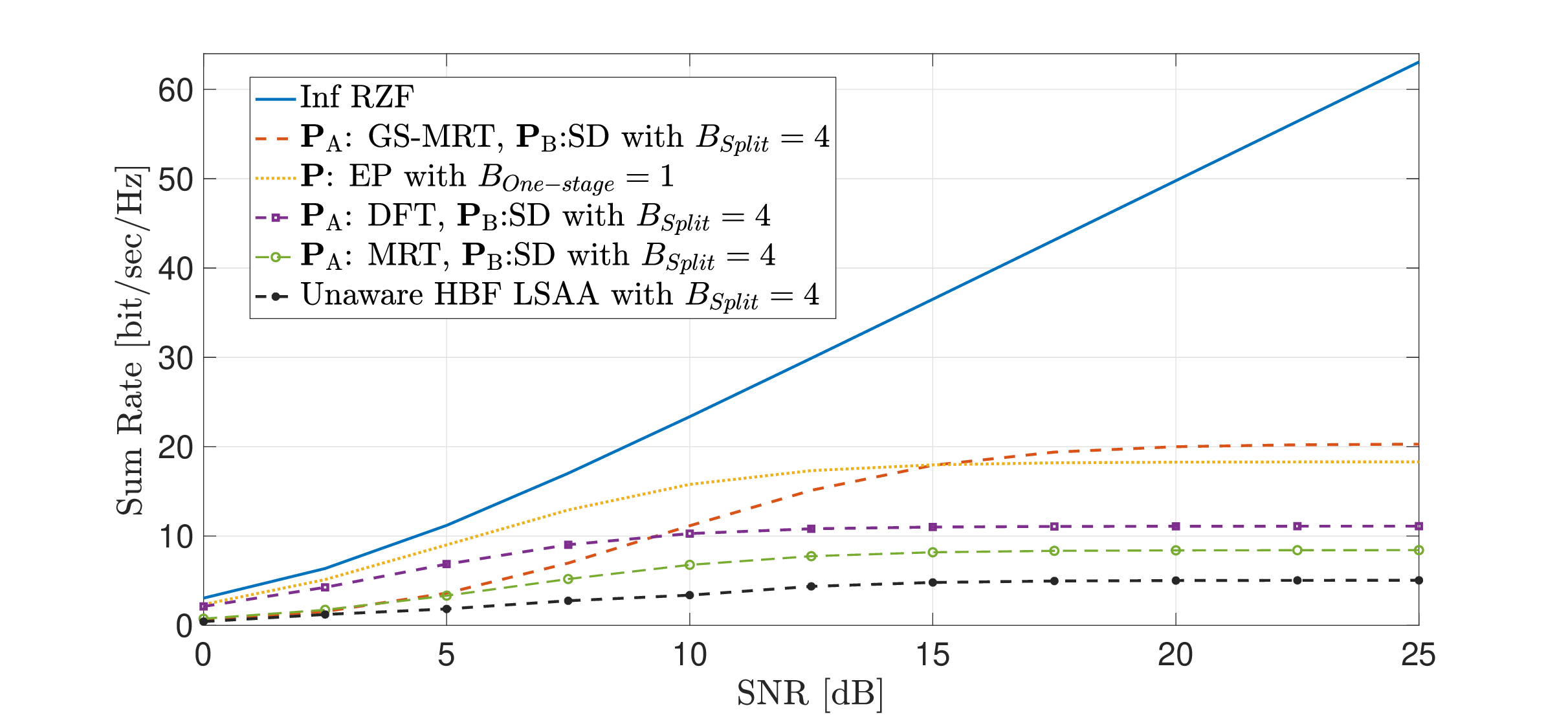} }} 
    \caption{The average sum rate versus the SNR for different precoding schemes.}
    \label{fig:all}
    \vspace{-5mm}
\end{figure}
{\section{Numerical Results}
In this section, we evaluate the performance of the proposed splitting precoding scheme and compare it with one-stage precoding, including the infinite-resolution (Inf) case using RZF precoding and the quantization-aware one-stage SESD baseline \cite{khorsandmanesh2023optimized}. The sum rate is adopted as the performance metric, since it offers a more intuitive measure of communication efficiency than the sum MSE used in the optimization. The average sum rate is computed as
\begin{equation}  \sum_{k=1}^K 
\mathbb{E} \left[\log_2 \left(1+\frac{\big|[{\mathbf{H}} \mathbf{P}]_{k,k}\big|^2}{\sum_{i=1, i  \ne k}^{K}\big|[{\mathbf{H}} \mathbf{P}
]_{k,i}\big|^2 +\sigma_0^2} \right)\right],\label{eq:sumrateaverage}
\end{equation}
where $\mathbf{P}$ denotes the precoding matrix obtained by each method, and the expectation is taken over the channel realizations using Monte Carlo simulations. We consider a normalized scenario in which all UEs experience the same large-scale fading coefficient $\gamma$. The receiver is assumed to have perfect CSI. In the following, we consider two channel models: i.i.d. Rayleigh fading and mmWave channels.

Unless stated otherwise, the baseline configuration consists of a BS equipped with $M=32$ antennas serving $K=8$ UEs with fronthaul bit resolution $B_{\text{Split}}=4$ and $N=K$. As stated in Remark 1, the resolution for the one-stage precoding for this setup should be set to $B_{\text{one-stage}}=1$ \cite{khorsandmanesh2023optimized}. All UEs experience a common SNR defined as $\text{SNR}=\frac{q \gamma}{\sigma_0^2}$ for a single antenna and single UE. 

Fig.~\ref{fig:all} illustrates the average sum rate as a function of the SNR for different precoding schemes. In Fig.~\ref{fig:compare}, the channel matrix~$\mathbf{H}$ follows an i.i.d. Rayleigh fading model. The “Inf RZF’’ curve exhibits an almost linear growth with SNR (in dB) since it is unquantized. Among the practical schemes, GS–MRT achieves the highest sum rate at high SNR. Its behavior follows that of MRT at low SNR and then saturates due to residual inter-user interference. The one-stage SESD baseline \cite{khorsandmanesh2023optimized} attains a similar slope but remains below the GS–MRT split architecture for the same fronthaul capacity at high SNR. At low SNR, there is a tradeoff between one-stage SESD and GS–MRT, as the splitting method has lower computational complexity, while one-stage SESD achieves a higher sum rate. The MRT and DFT AAS precoding achieve substantially lower rates and saturate early, reflecting their limited interference suppression capability. Due to fronthaul constraints, the BBU precoding cannot eliminate all interference; hence, the AAS precoding must select near-orthogonal subspaces for robust performance. In i.i.d. Rayleigh fading, the DFT structure does not provide sufficiently orthogonal beams, which explains the degradation in its performance. Overall, these results show that while simple splitting schemes such as MRT and DFT remain attractive due to their low complexity, quantization-aware or structure-optimized designs (e.g., SESD one-stage or GS–MRT splitting) provide substantial gains, especially at high SNR. Fig.~\ref{fig:mmwave} extends the comparison to a mmWave channel using a modified Saleh–Valenzuela model with $4$-tap Rician fading channel with a 10 dB Rician factor and $64$ subcarriers. As $M$ increases, the complexity of one-stage becomes prohibitive; therefore, for this baseline, we employ the expectation propagation (EP) method \cite{wang2020expectation}. The overall performance trends remain consistent with the Rayleigh case, with one notable exception: under mmWave channels, DFT outperforms MRT at high SNR due to its natural alignment with sparse angular structures. We also include the “unaware HBF LSAA’’ benchmark \cite{sohrabi2017hybrid}, where the baseband precoding in a hybrid beamforming setup is quantized without accounting for quantization effects. The clear loss observed in this curve highlights the importance of quantization-aware precoding at the BBU.

Fig.~\ref{fig:dft} shows the average sum rate versus SNR for different system configurations under DFT-based split precoding. The channel is i.i.d. Rayleigh fading. The results show that performance is strongly influenced by the BBU precoding dimension~$N$, and the quantization resolution~$B_{\rm Split}$. Increasing both the dimension and quantization resolution (e.g., $N=2K$ and $B_{\rm Split}=4$) yields a substantial rate improvement, approaching the performance of the proposed \emph{GS–MRT}. In contrast, configurations with $B_{\rm Split}=1$ saturate rapidly, indicating that coarse quantization severely limits multi-user interference suppression at high SNR. The case $N=M$ with $B_{\rm Split}=1$ achieves a moderate rate and behaves similarly to the \emph{SESD one-stage} baseline, illustrating the tradeoff between design simplicity and quantization precision. For a fixed~$B_{\rm Split}$, doubling from $N=K$ to $N=2K$ provides a clear gain in the low-to-medium SNR regime, but saturation remains at high SNR since quantization-induced distortion eventually dominates. This figure highlights that jointly increasing the effective channel dimension and quantization resolution is essential for unlocking higher throughput, whereas improving only one of the two parameters yields diminishing returns under tight fronthaul constraints. From a computational perspective, the one-stage architecture with SESD-based precoding at the BBU requires $O(KL^{2\gamma M})$ operations for $0 \leq \gamma \leq 1$~\cite{khorsandmanesh2023optimized}. In the split architecture, the BBU complexity reduces to $O(KL^{2\gamma N})$, which, although still exponential, is significantly lower for small~$N$. The AAS-side complexity depends on the subspace selection method: GS–MRT requires $O(MN^2)$, while a DFT-based implementation using an FFT reduces the cost to $O(M \log M)$ and is executed only once. In summary, MRT and DFT offer the lowest complexity but suffer from noticeable performance loss, while the GS–MRT split design provides an attractive balance between computational efficiency and achievable rate.
\vspace{-4mm}
\section{Conclusion}
This letter introduced a quantization-aware splitting precoding for fronthaul-limited downlink massive MIMO systems. The AAS performs the first precoding stage by selecting the relevant subspaces, while the BBU refines the transmission using quantization-aware MMSE precoding, which is solved via SESD. The simulation results demonstrate that the proposed design outperforms the conventional one-stage approach at high SNR when both operate under comparable fronthaul resolution constraints. The GS-MRT and SESD one-stage baselines achieve near-ideal performance, whereas the DFT and MRT schemes constitute practical low-complexity alternatives. For the DFT-based method, increasing the BBU precoding dimension and quantization resolution leads to further sum-rate gains. Overall, the proposed architecture offers an attractive tradeoff between fronthaul efficiency and computational complexity, establishing a scalable and quantization-resilient solution for massive MIMO deployments. 

\begin{figure}[!t]
        \centering
        \includegraphics[width=1\linewidth]{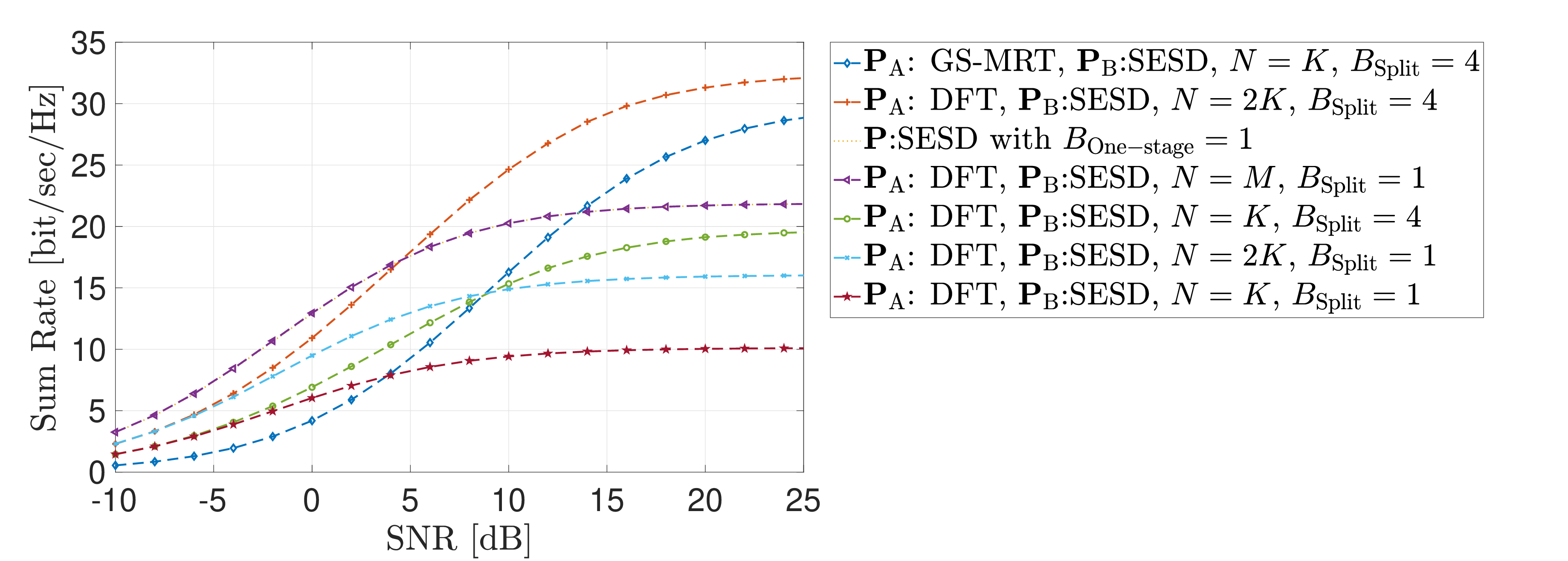}
        \caption{The average sum rate versus the SNR for different setups with $M=32$ and $K=8$.}
        \label{fig:dft}
        \vspace{-3mm}
\end{figure} 
\bibliographystyle{IEEEtran}
\bibliography{IEEEabrv,references}

\end{document}